\documentclass[pra,aps,twocolumn]{revtex4-1}
\usepackage{amsmath}
\usepackage{amssymb}
\usepackage{graphicx}
\newcommand{\mP}{{\mathcal P}}
\newcommand{\mT}{{\mathcal T}}

\newcommand{\beqa}{\begin{eqnarray}}
\newcommand{\eeqa}{\end{eqnarray}}

\begin{document}
\title{Origin of maximal symmetry breaking in even $\mP\mT$-symmetric lattices} 
\author{Yogesh N. Joglekar}
\email{yojoglek@iupui.edu}
\author{Jacob L. Barnett}
\affiliation{Department of Physics, 
Indiana University Purdue University Indianapolis (IUPUI), 
Indianapolis, Indiana 46202, USA}
\date{\today}
\begin{abstract}
By investigating a parity and time-reversal ($\mP\mT$) symmetric, $N$-site lattice with impurities $\pm i\gamma$ and hopping amplitudes $t_0 (t_b)$ for regions outside (between) the impurity locations, we probe the origin of maximal $\mP\mT$-symmetry breaking that occurs when the impurities are nearest neighbors. Through a simple and exact derivation, we prove that the critical impurity strength is equal to the hopping amplitude between the impurities, $\gamma_c=t_b$, and the simultaneous emergence of $N$ complex eigenvalues is a robust feature of any $\mP\mT$-symmetric hopping profile. Our results show that the threshold strength $\gamma_c$ can be widely tuned by a small change in the global profile of the lattice, and thus have experimental implications. 
\end{abstract}
\maketitle
%---------------------------------------------------------------------------------%

\noindent{\it Introduction:} The discovery of ``complex extension of quantum mechanics'' by Bender and coworkers~\cite{bender1,bender2} set in motion extensive mathematical~\cite{mostafa1,mostafa2,mostafa3} and theoretical investigations~\cite{reviews} of non-Hermitian Hamiltonians $H_{\mP\mT}=\hat{K}+\hat{V}$ that are symmetric with respect to combined parity ($\mP$) and time-reversal ($\mT$) operations. Such continuum or lattice Hamiltonians~\cite{znojil1,znojil2,znojil3,spin}  usually consist of a Hermitian kinetic energy part, $\hat{K}=\hat{K}^\dagger$, and a non-Hermitian, $\mP\mT$-symmetric potential part, $\hat{V}=\mP\mT\hat{V} \mP\mT\neq\hat{V}^\dagger$. Although it is not Hermitian $H_{\mP\mT}$ has purely real eigenvalues $E=E^*$  over a range of parameters, and its eigenfunctions are simultaneous eigenfunctions of the combined $\mP\mT$-operation; this range is defined as the $\mP\mT$-symmetric region. The breaking of $\mP\mT$-symmetry, along with the attendant non-reciprocal behavior, was recently observed in two coupled optical waveguides~\cite{expt1,expt2} and has ignited further interest in $\mP\mT$-symmetric lattice models. These evanescently coupled waveguides provide an excellent realization~\cite{christo} of an ideal, one-dimensional lattice with tunable hopping~\cite{gf}, disorder~\cite{berg1}, and non-Hermitian, on-site, impurity potentials~\cite{bendix,song}. 

Recently nonuniform lattices with site-dependent hopping $t_\alpha(k)=t_0\left[k(N-k)\right]^{\alpha/2}$ and a pair of imaginary impurities $\pm i\gamma$ at positions $(m,\bar{m})$ have been extensively explored~\cite{song,longhi,clint,derek}, where $\bar{m}=N+1-m$ and $N\gg 1$ is the number of lattice sites. The $\mP\mT$-symmetric phase in such a lattice is robust when $\alpha\geq 0$, the loss and gain impurities $\pm i\gamma$ are closest to each other, and $\gamma\leq \gamma_c$ where the critical impurity strength is proportional to the bandwidth of the clean lattice, $\gamma_c\propto 4t_0(N/2)^\alpha$. For a generic impurity position $m$, when the impurity strength $\gamma>\gamma_c(m)$ increases the number of complex eigenvalues increases sequentially from four to $N-1$ when $N$ is odd and to $N$ when it is even. {\it In an exceptional contrast}, when $m=N/2$ - nearest neighbor impurities on an even lattice - all eigenvalues simultaneously become complex at the onset of $\mP\mT$-symmetry breaking. This maximal symmetry breaking is accompanied by unique signatures in the time-evolution of a wavepacket~\cite{derek}. 

These results raise the following questions: Is this exceptional behavior limited to lattices with $\alpha$-dependent hopping or is it generic? Which factors truly determine the critical impurity strength $\gamma_c(N/2)$ in the exceptional case? How does the critical impurity strength $\gamma_c(m)$ depend upon lattice parameters and impurity positions? 

In this Brief Report, we investigate an $N$-site lattice with impurities $\pm i\gamma$ at positions $(m,\bar{m})$ and a constant hopping amplitude $t_0 (t_b)$ for sites outside (between) the parity-symmetric impurity locations. Our two salient results are as follows: i) When $m=N/2$, we analytically prove that all eigenvalues simultaneously become complex when $\gamma>\gamma_c(N/2)=t_b$. This robust result is true for {\it any symmetric distribution of real hopping amplitudes}.  ii) When $t_b\gg t_0$, the critical impurity strength $\gamma_c(m) \rightarrow t_b$ irrespective of the impurity position $m$. When $t_b< t_0$, the critical impurity strength $\gamma_c(m)\sim t_b^{\eta}$ where the exponent $\eta(d)\sim d$ increases monotonically with the distance $d=N+1-2m$ between the impurities. Thus, the $\mP\mT$-symmetry breaking threshold can be substantially tuned without significant changes in the global hopping-amplitude profile of the lattice, and the exceptional nature of the $m=N/2$ case is due to the ability to partition the system into two, and exactly two, pieces. 

%---------------------------------------------------------------------------------%

\noindent{\it Tight-binding Model:} We start with the Hamiltonian for a one-dimensional, tight-binding, non-uniform lattice 
\begin{equation}
\label{eq:h}
H_{\mP\mT}=-\sum_{i=1}^{N-1} t(i)\left(a^{\dagger}_{i+1} a_i + a^{\dagger}_i a_{i+1}\right) + i\gamma\left(a^{\dagger}_{m}a_{m}-a^{\dagger}_{\bar{m}}a_{\bar{m}}\right), 
\end{equation}
where $a^{\dagger}_n (a_n)$ is the creation (annihilation) operator for a state localized at site $n$, 
and the hopping function is given by $t(i)=t_b>0$ for $m\leq i\leq \bar{m}-1$, and $t(i)=t_0>0$ otherwise. This Hamiltonian continuously extrapolates from that for a lattice of length $d=N+1-2m$ with impurities at its end when $t_b\gg t_0$, to that of a pair of disconnected lattices, one with the gain impurity and the other with the loss impurity, when $t_b\ll t_0$. Note that the critical impurity strengths in these two limits are known~\cite{song,mark}. Due to the constant hopping amplitude outside or between the impurity locations, an arbitrary eigenfunction $|\psi\rangle=\sum_{n=1}^{N}\psi(n) a^{\dagger}_n|0\rangle$ with energy $E$ can be expressed using the Bethe {\it ansatz} as 
\begin{equation}
\psi(n)=\left\{
\begin{array}{cc}
A\sin(kn), & 1\leq n\leq m,\\
P\sin(k'n)+Q\cos(k'n), & m< n< \bar{m},\\
B\sin(k\bar{n}), & \bar{m}\leq n\leq N.\\
\end{array}\right.
\label{eq:bethe}
\end{equation}
Here $E(k,k')=-2t_0\cos(k)=-2t_b\cos(k')$ defines the relation between the quasimomenta $k,k'$. In the $\mP\mT$-symmetric phase, the energy spectrum of Eq.(\ref{eq:h}) is particle-hole symmetric~\cite{mapping}, and the eigenenergies satisfy $|E|\lesssim 2\max(t_0,t_b)$. Note that the relative phases of $\psi(n)$ are the same at different points within each of the three regions, although there may be a phase difference between wavefunctions in different regions. Therefore, without loss of generality, {\it we may choose $\psi(n)$ to be real for $1\leq n\leq m$}. By considering the eigenvalue equation $H_{\mP\mT}|\psi\rangle=E|\psi\rangle$ at points $m,m+1$ and their reflection counterparts, it follows that the quasimomenta $(k,k')$ obey the equation~\cite{mark}
\begin{eqnarray}
M(k,k') &\equiv \left[\sin^2\left[k(m+1)\right]+\Gamma^2\sin^2(km)\right]\nonumber\\
& \times \sin\left[k'(N+1-2m)\right] +T_b^2\sin^{2}(km)\nonumber\\
& \times \sin\left[k'(N-1-2m)\right] -2T_b\sin(km)\nonumber\\
& \times \sin\left[k(m+1)\right]\sin\left[k'(N-2m)\right]=0,
\label{eq:quasi}
\end{eqnarray}
where $\Gamma=\gamma/t_0$ and $T_b=t_b/t_0$ denote the dimensionless impurity strength and hopping amplitude respectively. Note that when $2\min(t_0,t_b)<|E|\leq 2\max(t_0,t_b)$, $k$ is real and $k'$ is purely imaginary (or vice versa), whereas for $|E|\leq 2\min(t_0,t_b)$, both $k,k'$ are real. Thus, Eq.(\ref{eq:quasi}) represents two distinct equations in these two cases. 

% phase diagram and scaling of the critical impurity strength.
\begin{figure}[htb]
\begin{center}
\includegraphics[angle=0,width=1.05\columnwidth]{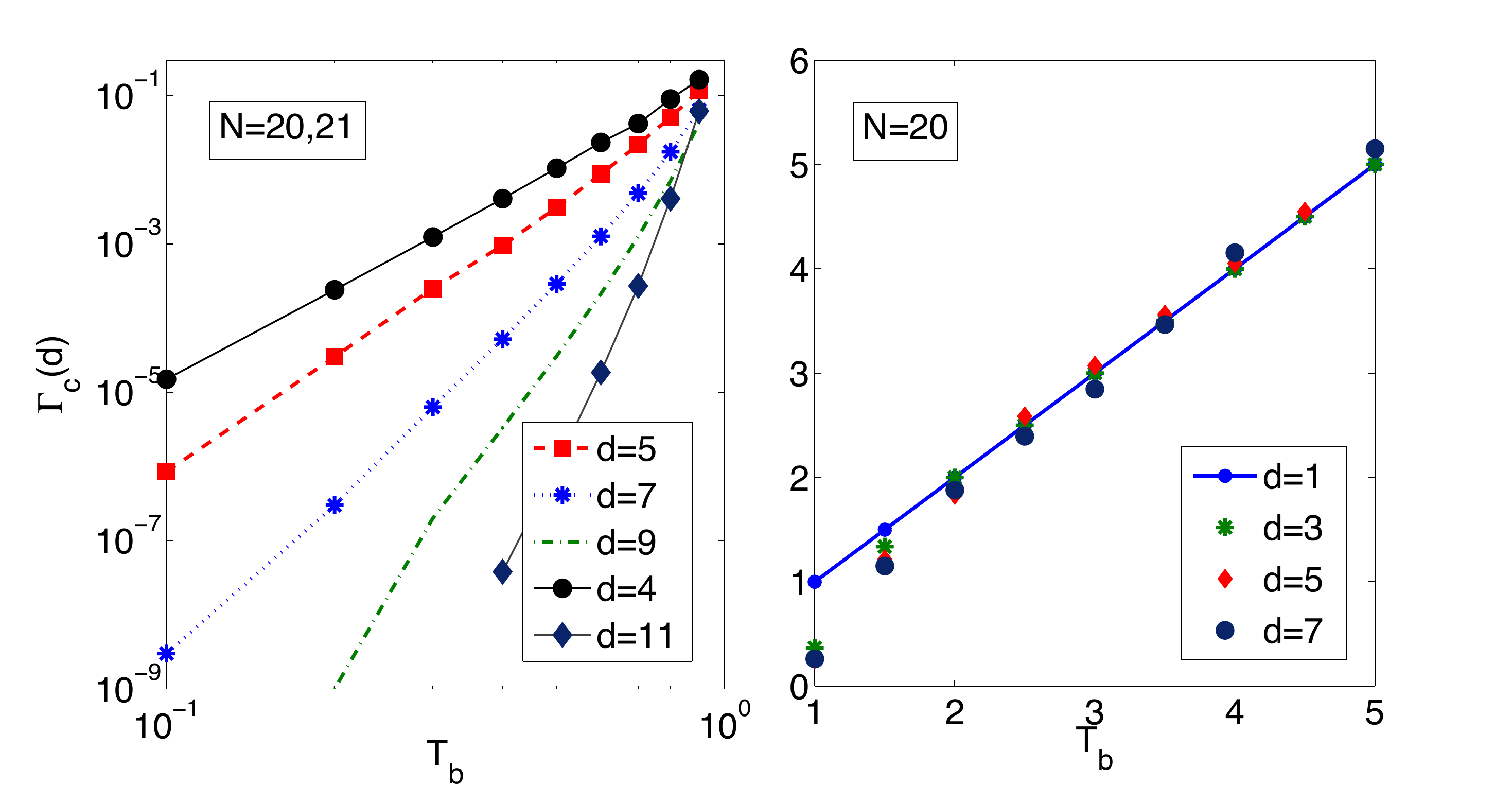}
\caption{(color online) a) Left-hand panel shows dimensionless critical impurity strength $\Gamma_c(d)=\gamma_c/t_0$ as a function of dimensionless hopping amplitude $0<T_b=t_b/t_0<1$ for various inter-impurity-distances $d$ in $N=20, 21$ lattices; note the logarithmic scale. It follows that $\Gamma_c(d)$ vanishes with a power-law behavior as $T_b\rightarrow 0$, as expected on physical grounds. b) Right-hand panel shows the critical impurity strength $\Gamma_c(d)$ as a function of $T_b\geq 1$ for various values of $d$.  Although at $T_b=1$, the critical strength $\Gamma_c(d)$ reduces with distance $d$ between the impurities, for $T_b\geq 2$ the critical impurity strength $\Gamma_c\rightarrow T_b$ ($\gamma_c\rightarrow t_b$) irrespective of $d$.}
\label{fig:phase}
\end{center}
\end{figure} 
The right-hand panel in Fig.~\ref{fig:phase} shows the dimensionless critical impurity strength $\Gamma_c(d)=\gamma_c(m)/t_0$ as a function of $T_b=t_b/t_0\geq 1$ for various inter-impurity-distances $d=N+1-2m$ in an $N=20$ even lattice; we obtain similar results for an odd lattice. We find that $\gamma_c\rightarrow t_b$ quickly for $t_b/t_0> 1$; when $t_b/t_0\gg1$, the lattice reduces to one with $d+1$ sites, impurities at its end points, and the result $\gamma_c=t_b$ is expected~\cite{mark}. The left-hand panel shows $\Gamma_c(d)$ vs. $T_b$ on a logarithmic scale in $N=20$ and $N=21$ lattices for $T_b<1$. As the distance $d$ between the impurities increases, corresponding critical impurity strength decreases as a power-law, $\Gamma_c(d)\propto T_b^{\eta(d)}$ where the exponent $\eta(d)\sim d$. This behavior can be qualitatively understood as follows: the system is in the $\mP\mT$-symmetric region if the frequency $\sim\gamma/t_0$ at which particles are created at the gain-impurity site $m$ is lower than rate at which these excess particles can hop over to the loss-impurity site, where they are absorbed at frequency $\sim\gamma/t_0$. Since $t_b$ is the hopping amplitude at sites between the impurities, it follows that the effective frequency of hopping from the gain- to the loss-site decreases with $d$ as $T_b^d$. Indeed, when $t_b/t_0\ll 1$, the system is divided into two, non $\mP\mT$-symmetric, uniform lattices, one with the loss impurity and the other with the gain. It follows, then, that $\gamma_c\rightarrow 0$ as $t_b/t_0\rightarrow 0$. 

%---------------------------------------------------------------------------------%

\noindent{\it Origin of Maximal Symmetry Breaking:} Now let us consider the $m=N/2$ case, where Eq.(\ref{eq:quasi}) reduces to
\begin{equation}
\label{eq:nn}
t_0^2\sin^2\left[k\left(\frac{N}{2}+1\right)\right]=\left(t_b^2-\gamma^2\right)\sin^2\left(\frac{kN}{2}\right).
\end{equation}
It follows from Eq.(\ref{eq:nn}) that the $\mP\mT$-symmetry breaks maximally when $\gamma>\gamma_c(N/2)=t_b$ and is accompanied by the simultaneous emergence of $N$ {\it complex} (not purely imaginary) quasimomenta and eigenenergies. Since the bandwidth of the clean lattice is determined by both hoppings $(t_0,t_b)$, it follows that the {\it critical impurity strength is independent of the lattice bandwidth}. 

To generalize this result, we consider the system with an arbitrary, $\mP\mT$-symmetric, position-dependent hopping profile  $t_k=t_{N-k}$ and real energy eigenvalues. Since the hopping and eigenvalues are real, the eigenvalue difference equations imply that for any eigenfunction $|\psi\rangle$, we can choose the coefficients $\psi(k)$ to be real for $1\leq k\leq m$. A real eigenvalue $\epsilon$ and the (real) coefficients $\alpha=\phi(N/2)$ and $\beta=\phi(N/2-1)$ of its corresponding eigenfunction $|\phi_\epsilon\rangle=\sum_{i=1}^{N}\phi(i)|i\rangle$ satisfy  
\begin{equation}
\det
\left[\begin{array}{cc}
t_{N/2-1}\beta+ (\epsilon-i\gamma)\alpha & t_{N/2}\alpha\\
t_{N/2}\alpha & t_{N/2-1}\beta+ (\epsilon+i\gamma)\alpha
\end{array}\right]=0,
\label{eq:det}
\end{equation}
where we have used the $\mP\mT$-symmetric nature of eigenfunctions to deduce that $\phi(N/2+1)=e^{i\chi}\alpha$, $\phi(N/2+2)=e^{i\chi}\beta$. Thus, when $\gamma>\gamma_c= t_{N/2}=t_b$, the eigenvalue $\epsilon$ must become complex. Since this result is true for all eigenfunctions, it follows that the $\mP\mT$-symmetry breaks maximally and the critical impurity strength is solely determined by the hopping amplitude  between the two impurities. This robust result also explains the fragile nature of $\mP\mT$-symmetric phase in lattices with hopping function $t_\alpha(k)$ for $\alpha<0$~\cite{derek}: in this case, the lattice bandwidth $\Delta_\alpha\sim N^{-|\alpha|/2}$ whereas the hopping amplitude between the two nearest-neighbor impurities scales as $t_b\sim N^{-|\alpha|}$. Therefore the critical impurity strength $\gamma_c/\Delta_\alpha\sim N^{-|\alpha|/2}\rightarrow 0$ as $N\rightarrow\infty$. A similar analysis for closest impurities in an odd-$N$ lattice shows that, due to the presence of a lattice site between the two impurity positions $m=(N-1)/2$ and $\bar{m}=(N+3)/2$, the corresponding critical impurity strength $\gamma_c$ depends on the details of the eigenfunction. 

Thus, the maximal symmetry breaking only occurs in an even, $\mP\mT$-symmetric lattice with nearest-neighbor impurities, and its origin is the ability to naturally partition such a lattice into exactly two components. 
%---------------------------------------------------------------------------------%

\noindent{\it Acknowledgments:} This work was supported by the IUPUI Undergraduate Research Opportunities Program (J.B.) and NSF Grant No. DMR-1054020 (Y.J.). 
 
%---------------------------------------------------------------------------------%

%---------------------------------------------------------------------------------%
\end{document}